\documentclass[aps,prl,twocolumn,superscriptaddress, showpacs]{revtex4}

\usepackage{graphicx}

\begin{document}

%\preprint{}

\title{Magnetic trapping of atomic nitrogen ($^{14}$N) and cotrapping of NH ($X^3\Sigma^-$)}


\author{Matthew T. Hummon}
\email{matt@cua.harvard.edu}
%\thanks{}
\affiliation{Department of Physics, Harvard University, Cambridge, MA 02138}
\affiliation{Harvard-MIT Center for Ultracold Atoms, Cambridge, MA 02138}

\author{Wesley C. Campbell}
\affiliation{Department of Physics, Harvard University, Cambridge, MA 02138}
\affiliation{Harvard-MIT Center for Ultracold Atoms, Cambridge, MA 02138}

\author{Hsin-I Lu}
\affiliation{School of Engineering and Applied Sciences, Harvard University, Cambridge, MA 02138}
\affiliation{Harvard-MIT Center for Ultracold Atoms, Cambridge, MA 02138}

\author{Edem Tsikata}
\affiliation{Department of Physics, Harvard University, Cambridge, MA 02138}
\affiliation{Harvard-MIT Center for Ultracold Atoms, Cambridge, MA 02138}

\author{Yihua Wang}
\affiliation{Department of Physics, Harvard University, Cambridge, MA 02138}
\affiliation{Harvard-MIT Center for Ultracold Atoms, Cambridge, MA 02138}

\author{John M. Doyle}
\affiliation{Department of Physics, Harvard University, Cambridge, MA 02138}
\affiliation{Harvard-MIT Center for Ultracold Atoms, Cambridge, MA 02138}


\date{\today}

\begin{abstract}
We observe magnetic trapping of atomic nitrogen ($^{14}$N) and cotrapping of ground state imidogen ($^{14}$NH, $X^3\Sigma^-$).  Both are loaded directly from a room temperature beam via buffer gas cooling.   We trap approximately $1 \times 10^{11}$ $^{14}$N atoms at a peak density of $5 \times 10^{11}$ cm$^{-3}$ at 550 mK.  The 12~+5/$-3$ s $1/e$ lifetime of atomic nitrogen in the trap is limited by elastic collisions with the helium buffer gas.  Cotrapping of $^{14}$N and $^{14}$NH is accomplished, with  $10^{8}$ NH trapped molecules at a peak density of $10^{8}$ cm$^{-3}$.  We observe no spin relaxation of nitrogen in collisions with helium.
\end{abstract}

%\keywords{}

\maketitle

% References should be done using the \cite, \ref, and \label command
%\section{}
%\subsection{}
%\subsubsection{}

The rapidly growing diversity of cold and ultracold atomic and molecular systems has opened the way for new applications and for the study of physical systems with new interactions.   Recent achievements have brought closer the realization of quantum information systems based on cold polar molecules \cite{Demille_qc, Andre_qc_etal} as well as new approaches to searches for physics beyond the Standard Model using cold molecules \cite{Kozlov_moleculeEDM}. Recently, several new interacting systems have been studied: a Bose-Einstein condensate with dipolar interactions \cite{Lahaye_Cr_dipolar_etal}, near-threshold inelastic collisions in cold polar molecules \cite{Gilijamse_OH_Xe_etal}, and laser cooling and magnetic trapping of non-S-state atoms \cite{Hancox_rare-earth_etal, McClelland_Er_mot}.  However, the promise of these new systems goes even further than these accomplishments in condensed matter and atomic physics. One prominent example is in the realm of cold chemistry. Reactions in cold gas phase atom-molecule systems play a key role in  dense interstellar clouds  \cite{Edvardsson_N_OH, Akyilmaz_NO_depletion_etal}. If cold cotrapped atoms and molecules could be created at sufficient density, the parameters characterizing these reactions could be studied in the laboratory. Such studies may also lead to observation of novel chemical reaction pathways such as tunneling \cite{Balakrishnan_chem, Krems_chem}.

To bring these ideas in quantum information, New Physics, and cold chemistry to fruition, collisions and interactions between cold and ultracold atoms and polar molecules must be understood. This requires high sample densities ($>10^{10}$ cm$^{-3}$). One promising route toward this is atom-molecule cotrapping. Currently, trapped polar molecules alone have not yet been produced at high enough densities to observe molecule-molecule collisions. Photoassociation of ultracold bialkali mixtures has produced magnetically and electrostatically trapped polar molecules at temperatures of 100s of $\mu$K and densities of $10^4$ to $10^5$ cm$^{-3}$ \cite{Wang_KRb_etal, Kleinert_twist}. Alternatively,  techniques for direct cooling and slowing of hot polar molecules, such as Stark deceleration \cite{Meerakker_Stark} and buffer gas cooling \cite{Weinstein_CaH_etal}, have produced trapped molecules at temperatures in the range of 10s to 100s of mK and with densities of up to $10^8$ cm$^{-3}$ \cite{Meerakker_Stark, Weinstein_CaH_etal, Sawyer_met_etal, Campbell_NH_vibration_etal}.  

Cotrapping could provide low temperatures and higher density via sympathetic cooling.  This is also a promising avenue toward the study of cold atom-molecule collisions. With this in mind, many are working to cotrap molecules with atoms, specifically laser cooled alkali-metal atoms \cite{Soldan_NH_Rb,Lara_OH_Rb_etal, Lara_Rb_OH_etal,Tacconi_NH_Rb, Schlunk_Rb_ac_etal}. An important alternative to that approach is buffer gas loading and magnetic cotrapping of non-alkali atoms. To achieve cooling of molecules in a cotrapped system, the atoms can be cooled evaporatively, and these in turn can cool the molecules through collisions.  If inelastic processes are limited, a rapid density increase would also occur.  The selection of the most advantageous (lowest inelastic loss rate) atom-molecule pair has been investigated in detail \cite{Soldan_NH_Rb,Lara_OH_Rb_etal, Lara_Rb_OH_etal,Tacconi_NH_Rb}. First, the molecule should be Hund's case (b), where the electron spin is weakly coupled to the internuclear axis. NH, a molecule studied by many groups \cite{Campbell_nh_trapping_etal,Hoekstra_NH_electro_etal, Lewandowski_NH}, is of this type. Second, the atomic partner should have low mass and low polarizability. Comparing the atom we have chosen, nitrogen (N), with a typical alkali metal atom, rubidium (Rb), N has $1/6$ the mass and $1/40$ the polarizability \cite{Lim_polarizability_etal, Stiehler_polarizabilites}. Calculations of cold collision properties for the $^{15}$N-$^{15}$N and $^{15}$N-$^{14}$N systems indicate favorable collision rates for atomic evaporative cooling \cite{Zygelman_dipolar}. Thus, this atom is set to be an excellent refrigerator, perhaps for a molecule like NH.
  
\begin{figure}[b]
 \includegraphics[width=86mm]{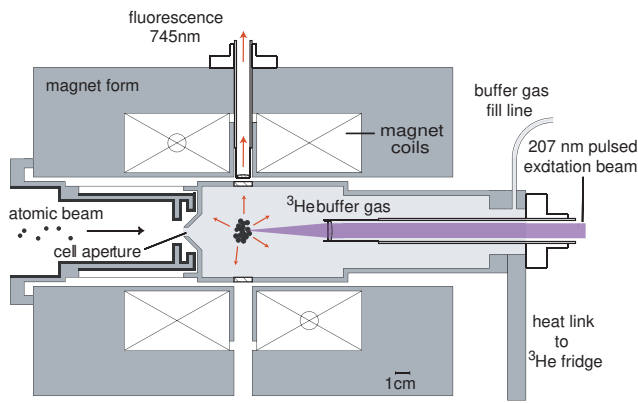}
 \caption{\label{fig_apparatus} Diagram of trapping apparatus.  The nitrogen atomic beam is produced by a radio frequency plasma source located 23 cm to the left of the cell aperture.  A lens is mounted inside the buffer gas cell 60 mm from the trap center to attain saturation of the two-photon transition for atomic nitrogen.  The trapping magnet surrounds the buffer gas cell and is described further in Ref. \cite{Campbell_nh_trapping_etal}.}
 \end{figure}

In this Letter we report the first trapping of atomic nitrogen ($^{14}$N) and cotrapping of $^{14}$N with ground state imidogen radicals $(^{14}$NH, $X^3\Sigma^-)$.  The apparatus, shown in Fig. \ref{fig_apparatus}, is similar to that described in Ref. \cite{Campbell_nh_trapping_etal}.  Briefly, a cylindrical copper buffer gas cell is thermally anchored to a $^3$He refrigerator and is filled with $^3$He buffer gas to a density of $(3$ to $10) \times 10^{14}$ cm$^{-3}$ at a temperature of about $550$ mK.  The buffer gas cell is surrounded by a pair of superconducting coils that produce a spherical quadrupole trapping field with maximum trap depth of $ 3.9$ T.  Cell windows are located both axially and radially along the cell to allow optical access for laser induced fluorescence spectroscopy of the trapped species.  The atoms and molecules are loaded into the buffer gas cell via a molecular beam that enters the cell through a 1 cm aperture at the edge of the trapping region.

To produce the high flux beam of atomic nitrogen we use an inductively coupled radio-frequency (RF) plasma source \cite{rf_source} operating in high brightness mode \cite{Moldovan_N_source_etal}.  The plasma source generates a flux of atomic nitrogen of about $10^{16}$ atoms\thinspace s$^{-1}$\thinspace sr$^{-1}$ \cite{Moldovan_N_source_etal}. The aperture of the plasma source is located about 23 cm from the entrance to the trapping cell.  As the plasma source is run continuously during the experiment, a shutter, heat sunk at a temperature of 77 K, is located between the plasma source and the trapping cell.  To load atoms or molecules into the cell, the shutter is pulsed opened for 20 ms and typically operates at a repetition rate of 4 Hz for continuous loading of the magnetic trap.

\begin{figure}
 \includegraphics[width=86mm]{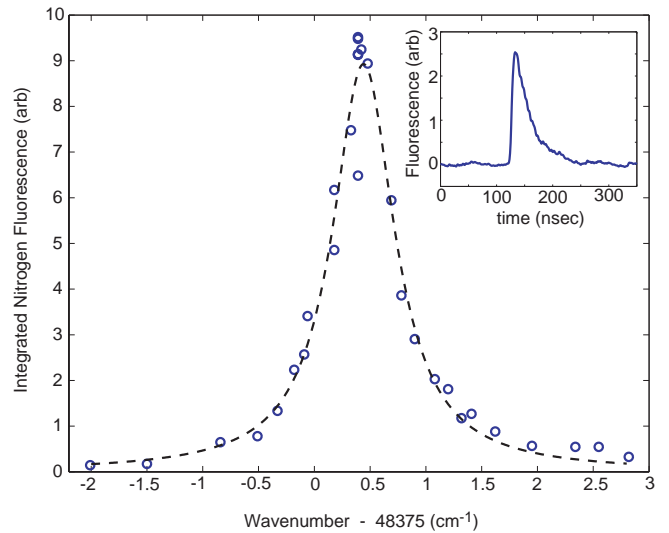}
 \caption{\label{fig_n_spectrum}Integrated nitrogen fluorescence vs. excitation wavenumber. The data (circles) is an average of 100 fluorescence acquisitions taken during continuous trap loading. The dashed line is a fit of the data to a Lorentzian.  The inset shows time resolved nitrogen fluorescence taken on resonance.}
 \end{figure}

For detection of magnetically trapped atomic nitrogen we use two-photon absorption laser induced fluorescence (TALIF) \cite{adams_TALIF}, as the most accessible single photon transition is at 120 nm \cite{nist_atomic_spectra_etal}.  We excite atomic nitrogen in the ground $(2p^3)^4$S$_{3/2}$ state with two photons at 207 nm to the excited $(3p)^4$S$_{3/2}$ state at 96750 cm$^{-1}$.  To generate the 207 nm light used for excitation we use third harmonic generation of 620 nm light generated by a nanosecond pulsed dye laser via nonlinear frequency conversions in KDP and BBO crystals. In order to generate the high intensities necessary to excite the two-photon transition, a  60 mm focal length lens is mounted in the buffer gas cell and used to focus $\approx 1 $ mJ of 207 nm light to a $1/e^2$ waist diameter of $\approx$ 40 $\mu$m at the trap center.  Using a lens mounted at the midplane of the trapping magnet, we collect the atomic fluorescence at $\approx 745$ nm from the decay of the excited $(3p)^4$S$_{3/2}$ state  to the $(3s)^4$P state and detect it using a photomultiplier tube (PMT).  The pulse duration of the excitation light is about 10 ns, and the lifetime of the excited  $(3p)^4$S$_{3/2}$ state is about 26 ns \cite{bengtsson_n_lifetime}, which allows us to detect a temporally resolved fluorescence signal from the atomic nitrogen, shown in the inset of Fig. \ref{fig_n_spectrum}.  Detection of NH is performed using laser induced fluorescence (LIF) excited on the $| A^3\Pi_2,\nu' = 0, N' = 1\rangle \leftarrow |X^3\Sigma^-, \nu'' = 0, N'' = 0\rangle$ transition.  The NH fluorescence is collected by the same lens that collects $^{14}$N fluorescence and, using a dichroic mirror, is sent to a separate PMT for detection.

Figure \ref{fig_n_spectrum} shows the TALIF spectrum of trapped $^{14}$N taken during continuous loading of the magnetic trap.  The spectrum fits well to a Lorentzian lineshape with a full-width half-maximum of 0.6 cm$^{-1}$.  The Doppler width and Zeeman broadening of the two-photon transition are negligible compared to the time averaged spectral width of the pulse laser, specified as 0.15 cm$^{-1}$. The experimental linewidth is consistent with power broadening.  Measurements of fluorescence signal versus excitation laser pulse energy and calculations of the two-photon excitation probability for the experimental parameters indicate that we achieve saturation of the TALIF transition.  Precise estimation of the peak density of atomic nitrogen in the trap using the TALIF measurements is challenging, as accurate knowledge of fluorescence collection efficiency, effective TALIF excitation volume, and fluorescence yield per laser pulse is required.  Furthermore, the pulsed dye laser generates multimode excitation light with a non-gaussian spatial profile and a spectral profile that changes from shot to shot.  Also, since a single 207 nm photon can photoionize the excited $(3p)^4$S$_{3/2}$ state of nitrogen, it is likely that a significant fraction (of order unity) of the excited nitrogen atoms undergo photoionization instead of fluorescence. In our data collection, we average the fluorescence over a large number of excitation pulses, typically about 100, so that the data yielded represents fluorescence taken over a range of saturation conditions.  To estimate the effective excitation volume, we use a theoretical value of the two-photon cross section \cite{Omidvar_cross_section, Omidvar_cross_section_err} and integrate the excitation probability over a gaussian spatial profile for the 207 nm light of $1/e^2$ diameter of 40 $\mu$m . Neglecting the effects of photoionization, the systematic uncertainties in the calculations of the fluorescence collection efficiency and effective excitation volume yields an overall uncertainty of a factor of 3 in the peak density  $n_{\text{N}} = 5 \times 10^{11}$ cm$^{-3}$ and total number $N_{\text{N}} =  1 \times 10^{11}$ of trapped  nitrogen atoms.   Including the effects of photoionization could increase the calculated number of trapped atoms by as much as a factor of 10 \cite{Bamford_O, Bell_PI_N}, so our determined $N_{\text{N}}$ may be considered a lower bound.

\begin{figure}
 \includegraphics[width=86mm]{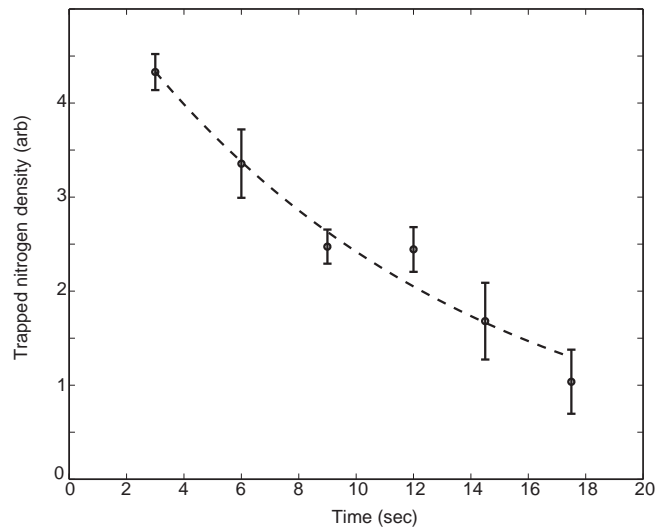}
 \caption{\label{fig_n_decay}Trapped nitrogen density vs. time after trap loading.  An exponential fit (dashed line) to the data yields a $1/e$ trap lifetime of 12+5/$-$3 s.}
 \end{figure}

Figure \ref{fig_n_decay} shows the integrated TALIF signal of the $^{14}$N in the magnetic trap versus time in the presence of $^3$He buffer gas of density $(6 \pm 1) \times 10^{14}$ cm$^{-3}$ and temperature $530 \pm 20$ mK.   As the trap depth for the nitrogen is $7$ K, many collisions of nitrogen with the helium are required to knock the nitrogen out of the trap, yielding a $1/e$ trap lifetime of 12 $+5$/$-3$ s.  In the absence of the trapping field, the diffusion lifetime of the nitrogen through the buffer gas would be on the order of 10 ms.  The 12 s lifetime is consistent with evaporative loss of nitrogen atoms due to elastic collisions with the helium buffer gas.  This puts a limit on the $^{14}$N-$^3$He inelastic collision rate coefficient of $k_{\text{in}} < 2.2 \times 10^{-16}$ cm$^3$\thinspace s$^{-1}$, corresponding to a limit of the $^{14}$N-$^3$He inelastic cross section of $\sigma_{\text{in}} < 3.3 \times 10^{-20}$ cm$^2$.  For comparison, the inelastic rate coefficient for the $^{52}$Cr-$^3$He system was measured to be $k_{\text{in}} = (2\pm1)\times 10^{-18}$ cm$^3$\thinspace s$^{-1}$ \cite{Weinstein_Cr}.  Non-S-state rare-earth atoms have inelastic rate coefficients with $^3$He in the range of $(2\times 10^{-16}< k_{\text{in}} < 7\times 10^{-15})$ cm$^3$\thinspace s$^{-1}$  \cite{Hancox_rare-earth_etal}.  Bismuth, though in the same group as nitrogen, has a large inelastic rate coefficient with $^3$He of $k_{\text{in}} > 1.8 \times 10^{-14}$ cm$^{3}$\thinspace s$^{-1}$ due to electronic anisotropy induced by spin-orbit coupling \cite{Maxwell_Bi_etal}.

\begin{figure}
 \includegraphics[width = 86mm]{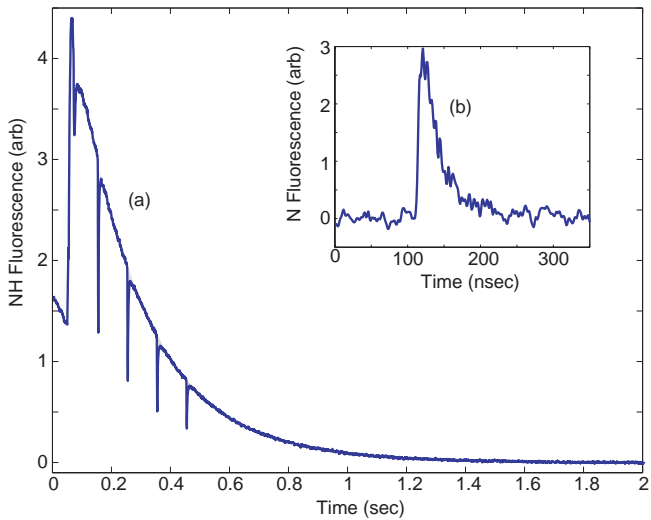}
 \caption{\label{fig_cotrap} Cotrapping of N and NH, (a) shows trapped NH with a 1/e lifetime of 250 ms. (b) shows the average atomic nitrogen fluorescence signal taken during the first 0.5 s of trace (a).  The downward spikes in trace (a) during the first 0.5 s correspond to the firing of the 207 nm excitation laser and the likely photodissociation of NH. }
 \end{figure}

To cotrap atomic nitrogen and NH radicals, we change the process gas for the RF plasma source.  We find that both ammonia or a combination of molecular nitrogen and hydrogen work equally well for producing the trap species.  Figure \ref{fig_cotrap} (a) shows the LIF signal of NH in the trap versus time. The NH has a $1/e$ trap lifetime of 250 ms, limited by Zeeman-state-changing collisions of NH with the helium buffer gas of density $(3\pm0.6) \times 10^{14}$~cm$^{-3}$ and temperature $550 \pm 20$ mK. The LIF signal has been calibrated using laser absorption \cite{Campbell_NH_vibration_etal}, and we trap about $10^8$ NH molecules at peak densities of $10^8$ cm$^{-3}$.  The downward spikes in the NH profile in  Fig. \ref{fig_cotrap} (a) correspond to the times at which the pulsed 207 nm laser fires for the atomic nitrogen detection. Figure \ref{fig_cotrap} (b) shows the average atomic nitrogen TALIF signal taken over the first 0.5 s of trapping, simultaneous with the trapping of NH shown in Fig \ref{fig_cotrap} (a).  The nitrogen fluorescence shown in Fig. \ref{fig_cotrap} (b) corresponds to about $4 \times 10^{10}$ trapped $^{14}$N atoms at a peak density of $1 \times 10^{11}$ cm$^{-3}$.  In the future, we should be able to increase the number of atoms and molecules loaded into the trap by at least a factor of 4 simply by moving the plasma source closer to the trapping cell.  The lifetimes of the cotrapped nitrogen and NH are long enough that thermal isolation via removal of the buffer gas should be possible \cite{Harris_valve_etal}, and is currently being pursued.  Nitrogen cotrapped with a molecule is an excellent candidate for sympathetic cooling.

As mentioned in the introduction to this paper, our work is important to understanding certain astrophysical phenomena. There has been considerable interest in interstellar nitrogen recently, as new astronomical observations have provided evidence that most interstellar nitrogen in the gas phase is not, as previously thought, in molecular form, but rather in atomic form \cite{Maret_N2_darkcloud, Knauth_interstellar_N2, Akyilmaz_NO_depletion_etal}.  A network of cold chemical reactions \cite{Sternberg_N_network} involving atomic nitrogen and neutral molecular radicals (OH, NO, CH, CN, and NH, in particular) \cite{Frankcombe_N_NH,Edvardsson_N_OH,Akyilmaz_NO_depletion_etal} play an important role in interstellar gas-phase chemistry models.  Specifically, the existence of small barriers ($\sim 25$ K) to these reactions is an important parameter and is beyond the sensitivity of current theoretical calculations \cite{Akyilmaz_NO_depletion_etal}.  Theoretical study of  these reactions predict rate constants on the order of $k \sim 10^{-11}$ cm$^3$\thinspace s$^{-1}$ at temperatures of $\sim 10$ K \cite{Edvardsson_N_OH, Frankcombe_N_NH}, suggesting that the reactions may be observable using a cold, high density sample of trapped atomic nitrogen, similar to what has been achieved here.  Removing the buffer gas, as we have done previously in similar work, would open the door to this measurement \cite{Harris_valve_etal}.

In summary,  we have observed magnetic trapping and cotrapping of atomic nitrogen and ground state NH molecules.   We trap approximately $1\times 10^{11}$ $^{14}$N atoms at a peak density of $5\times10^{11}$ cm$^{-3}$ and temperature of $550$ mK with a $1/e$ trap lifetime of $12$~+5/$-3$~s.   This lifetime sets a limit on the $^{14}$N-$^3$He inelastic collision rate coefficient of $k_{\text{in}} < 2.2 \times 10^{-16}$ cm$^3$\thinspace s$^{-1}$. Moreover, we cotrap about $4\times 10^{10}$ $^{14}$N atoms and $10^8$ NH molecules at a temperature of $550$ mK.  
If this system had the buffer gas removed \cite{Harris_valve_etal}, a wealth of research paths exist:  evaporative cooling, possibly to quantum degeneracy, of a new atomic species, investigation of sympathetic cooling of polar molecules in a magnetic trap, and cold nitrogen chemistry.

We would like to thank Gale Petrich and Leslie Kolodziejski for loan of the RF plasma source.  This work was supported by the National Science Foundation under Grant No. 0457047, the U.S. Department of Energy under Contract No. DE-FG02-02ER15316, and the U.S. Army Research Office.

% If in two-column mode, this environment will change to single-column
% format so that long equations can be displayed. Use
%\begin{widetext}
%\end{widetext}

% figures should be put into the text as floats.
% Use the graphics or graphicx packages (distributed with LaTeX2e)
% and the \includegraphics macro defined in those packages.
% See the LaTeX Graphics Companion by Michel Goosens, Sebastian Rahtz,
% entire page. There is no need to do explicit centering.

% \begin{figure}
% \end{figure}
\bibliography{nitrogen}

\end{document}